\documentclass[a4paper]{article}

\usepackage{INTERSPEECH2019}

\usepackage{url}
\usepackage{booktabs}
\usepackage{xcolor}
\usepackage{hyperref}
\newcommand{\coollink}[1]{\href{https://#1}{\nolinkurl{#1}}}
\usepackage{bm}

\title{Expediting TTS Synthesis with Adversarial Vocoding}
\name{\text{*}Paarth Neekhara$^1$, \text{*}Chris Donahue$^{2}$, Miller Puckette$^2$, Shlomo Dubnov$^2$, Julian McAuley$^1$}
\address{
  $^1$UC San Diego Department of Computer Science\\
  $^2$UC San Diego Department of Music\\
  \text{*} Equal contribution}
\email{pneekhar@eng.ucsd.edu, cdonahue@ucsd.edu}

\begin{document}

\maketitle
\begin{abstract}
Recent approaches in text-to-speech (TTS) synthesis employ neural network strategies to vocode perceptually-informed spectrogram representations directly into listenable waveforms. 
Such vocoding procedures create a computational bottleneck in modern TTS pipelines. 
We propose an alternative approach which utilizes generative adversarial networks (GANs) to learn mappings from perceptually-informed spectrograms to simple magnitude spectrograms which can be heuristically vocoded. 
Through a user study, we show that our approach significantly outperforms na\"ive vocoding strategies while being hundreds of times faster than neural network vocoders used in state-of-the-art TTS systems. 
We also show that our method can be used to achieve state-of-the-art results in unsupervised synthesis of individual words of speech.
\end{abstract}


\section{Introduction}

Generating natural-sounding speech from text is a well-studied problem with numerous potential applications. 
While past approaches were built on extensive engineering knowledge in the areas of linguistics and speech processing (see \cite{zen2009statistical} for a review), 
recent approaches adopt neural network strategies which learn from data to map linguistic representations into audio waveforms~\cite{arik2017deep,gibiansky2017deep,ping2017deep,wang2017tacotron,shen2018natural}. 
Of these recent systems, 
the best performing~\cite{ping2017deep,shen2018natural} are both comprised of two functional mechanisms which 
(1) map language into \emph{perceptually-informed spectrogram} representations (i.e.,~time-frequency decompositions of audio with logarithmic scaling of both frequency and amplitude), and 
(2) \emph{vocode} the resultant spectrograms into listenable waveforms. 
In such two-step TTS systems, 
using perceptually-informed spectrograms as intermediaries is observed to have empirical benefits over using representations which are simpler to convert to audio~\cite{ping2017deep}. 
Hence, vocoding is central to the success of state-of-the-art TTS systems, and is the focus of this work.

The need for vocoding arises from the non-invertibility of perceptually-informed spectrograms. 
These compact representations
exclude much of the information in an audio waveform, 
and thus require a predictive model to fill in the missing information needed to synthesize natural-sounding audio. 
Notably, standard spectrogram representations discard phase information resulting from the short-time Fourier transform (STFT), 
and additionally compress the linearly-scaled frequency axis of the STFT magnitude spectrogram into a logarithmically-scaled one. 
This gives rise to two corresponding vocoding subproblems: 
the well-known problem of \emph{phase estimation}, 
and the less-investigated problem of \emph{magnitude estimation}.

Vocoding methodology in state-of-the-art TTS systems~\cite{ping2017deep,shen2018natural} endeavors to address the joint of these two subproblems, 
i.e.,~to transform perceptually-informed spectrograms directly into waveforms. 
Specifically, both systems use WaveNet~\cite{oord2016wavenet} conditioned on spectrograms. 
This approach is problematic as it necessitates running WaveNet once per individual audio sample (e.g. $22050$ times per second), bottlenecking the overall TTS system as the language-to-spectrogram mechanisms are comparatively fast.\footnote{In our empirical experimentation with open-source codebases, the autoregressive vocoding phase was over $1500$ times slower on average than the language to spectrogram phase.}
Given that joint solutions currently necessitate such computational overhead, 
it may be methodologically advantageous to combine solutions to the individual subproblems.

Before endeavoring to develop individual solutions to magnitude and phase estimation, 
we first wished to discover which (if any) of the two represented a greater obstacle to vocoding. 
To answer this, we conducted a user study examining the effect that common heuristics for each subproblem have on the perceived naturalness of vocoded speech (Table~\ref{tab:gl}).\footnote{Sound examples: \coollink{chrisdonahue.com/advoc_examples}} 
Our study demonstrated that combining an ideal solution to \emph{either} magnitude or phase estimation with a heuristic for the other results in high-quality speech. 
Hence, 
we can focus our research efforts on \emph{either} subproblem, 
in the hopes of developing methods which are more computationally efficient than existing end-to-end strategies.

In this paper, 
we seek to address the magnitude estimation subproblem, 
which has received less attention in comparison to phase estimation~\cite{griffinlim,lws,li2016iterative,oyamada2018generative}.
We propose a learning-based method which uses Generative Adversarial Networks~\cite{goodfellow2014generative} to learn a stochastic mapping from perceptually-informed spectrograms into simple magnitude spectrograms. 
We combine this magnitude estimation method with a modern phase estimation heuristic, 
referring to this method as \emph{adversarial vocoding}.
We show that adversarial vocoding can be used to expedite TTS synthesis and additionally improves upon the state of the art in unsupervised generation of individual words of speech.

\subsection{Summary of contributions}

\begin{itemize}
    \item For both real spectrograms and synthetic ones from TTS systems, we demonstrate that our proposed vocoding method yields significantly higher mean opinion scores
    than a heuristic baseline and faster speeds than state-of-the-art vocoding methods.
    \item We show that our method can effectively vocode highly-compressed ($13$:$1$) audio feature representations.
    \item We show that our method improves the state of the art in unsupervised synthesis of individual words of speech.
    \item We measure the perceived effect of inverting the primary sources of compression in audio features. We observe that coupling solutions to either compression source with a heuristic for the other result in high-quality speech.
\end{itemize}

\section{Audio feature preliminaries}

\label{sec:feature}
\begin{figure}[t]
    \centering
    \includegraphics[width=1.\linewidth]{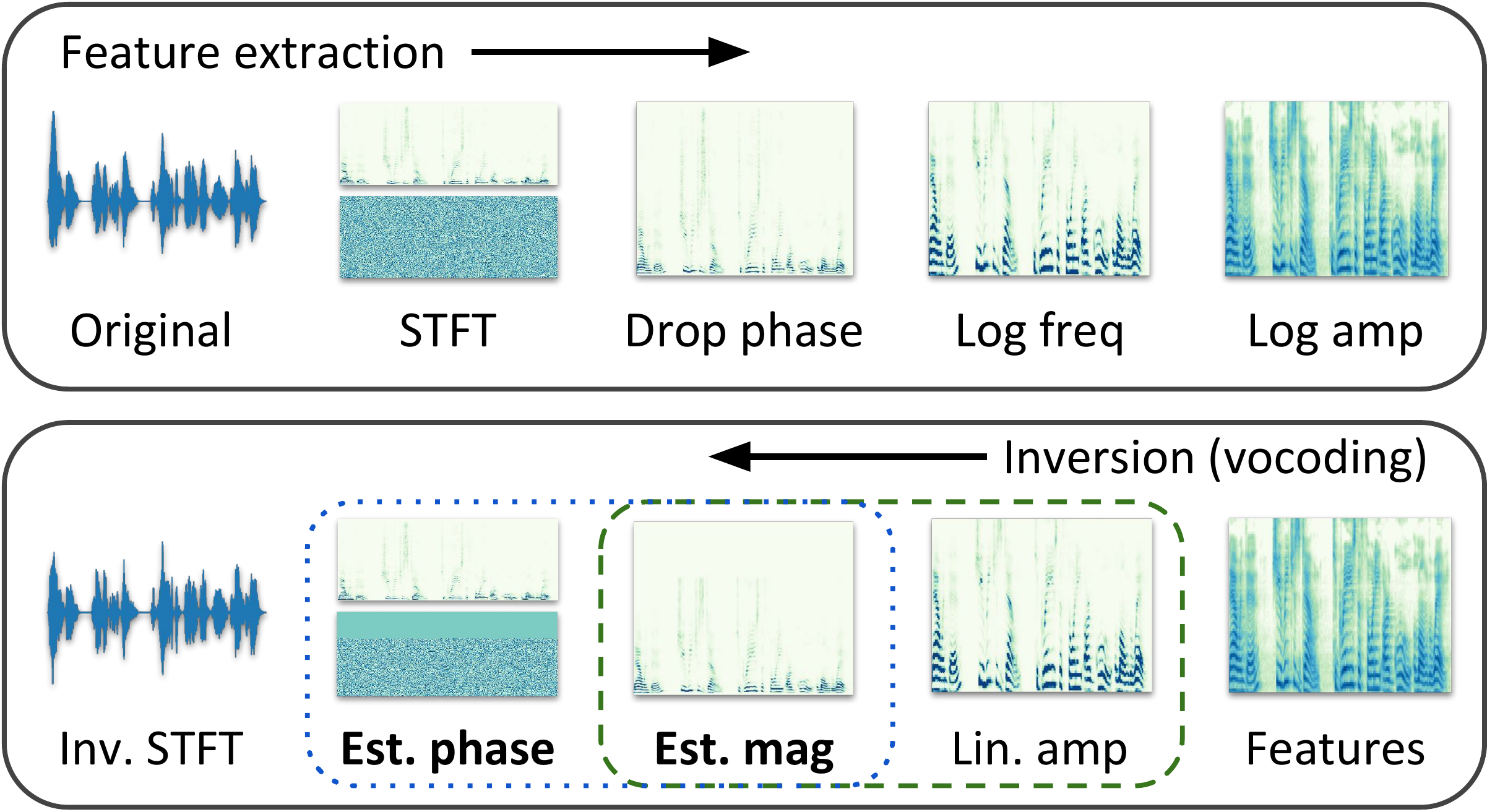}
    \caption{
    Depiction of stages in common audio feature extraction pipelines and corresponding inversion. 
    The two obstacles to vocoding are (1) estimating linear-frequency magnitude spectra from log-frequency mel spectra (outlined in \textcolor[rgb]{0.29, 0.33, 0.13}{green} dashed line), and (2) estimating phase information from magnitude spectra (outlined in \textcolor{blue}{blue} dotted line).
    We focus on magnitude estimation in this paper, observing that coupling an ideal solution to this subproblem with a phase estimation heuristic can produce high-quality speech (Table~\ref{tab:gl}).}
    \label{fig:extract_invert}
\end{figure}

The typical process of transforming waveforms into perceptually-informed spectrograms involves several cascading stages.
Here, we describe spectrogram methodology common to two state-of-the-art TTS systems~\cite{ping2017deep,shen2018natural}. A visual representation is shown in Figure~\ref{fig:extract_invert}.

\textbf{Extraction}~~~
The initial stage consists of decomposing waveforms into time and frequency using the STFT. Then,
the phase information is discarded from the complex STFT coefficients leaving only the linear-amplitude magnitude spectrogram. 
The linearly-spaced frequency bins of the resultant spectrogram are then compressed to fewer bins which are equally-spaced on a logarithmic scale (usually the mel scale~\cite{stevens1937scale}). 
Finally, amplitudes of the resultant spectrogram are made logarithmic to conform to human loudness perception, then optionally clipped and normalized.

\textbf{Inversion}~~~
To heuristically invert this procedure (vocode), 
the inverse of each cascading step is applied in reverse.
First, logarithmic amplitudes are converted to linear ones. 
Then,  
an appropriate magnitude spectrogram is estimated from the mel spectrogram.
Finally, appropriate phase information is estimated from the magnitude spectrogram, and the inverse STFT is used to render audio.

Unless otherwise specified, throughout this paper we operate on waveforms sampled at $22050$Hz using an STFT with a window size of $1024$ and a hop size of $256$. 
We compress magnitude spectrograms to $80$ bins ($\mathit{melBins}=80$) equally spaced along the mel scale from $125$Hz to $7600$Hz.
We apply log amplitude scaling and normalize resultant mel spectrograms to have $120$dB dynamic range. 
Precisely recreating this representation~\cite{mcfee2019open} is simple in our codebase.\footnote{Code: \coollink{github.com/paarthneekhara/advoc}}

\section{Measuring the effect of magnitude and phase estimation on speech naturalness}
\label{sec:magphs}

The audio feature extraction pipelines outlined in Section~\ref{sec:feature} have two sources of compression: the discarding of phase information and compression of magnitude information. 
Conventional wisdom suggests that the primary obstacle to inverting such features is phase estimation. 
However, 
to the best of our knowledge, 
a systematic evaluation of the individual contributions of magnitude and phase estimation on perceived naturalness of vocoded speech has never been reported.

To perform such an evaluation, we mix and match methods for estimating both STFT magnitudes and phases from log-amplitude mel spectrograms. 
A common heuristic for magnitude estimation is to project the mel-scale spectrogram onto the pseudoinverse of the mel basis which was originally used to generate it. 
As a phase estimation baseline, state-of-the-art TTS research~\cite{ping2017deep,shen2018natural} compares to the iterative Griffin-Lim~\cite{griffinlim} strategy with $60$ iterations. 
We additionally consider the more-recent Local Weighted Sums (LWS)~\cite{lws} strategy which, on our CPU, is about six times faster than $60$ iterations of Griffin-Lim.
As a proxy for an ideal solution to either subproblem, 
we also use magnitude and phase information extracted from real data. 

We show human judges the same waveform vocoded by six different magnitude and phase estimation combinations (inducing a comparison) and ask them to rate the naturalness of each on a subjective $1$ to $5$ scale (full user study methodology outlined in Section~\ref{sec:ljspeechexp}). 
Mean opinion scores are shown in Table~\ref{tab:gl}, and we encourage readers to listen to our sound examples linked from the footnote on the first page to help contextualize.

\begin{table}[t]
\centering
\caption{Ablating the effect of heuristics for magnitude and phase estimation on mean opinion score (MOS) of speech naturalness with $95$\% confidence intervals.
\textbf{Bolded} entries show that coupling an ideal solution to either subproblem (real data used as a proxy) with a good heuristic for the other yields speech with only $2$--$9$\% lower
MOS than real speech ($p<0.05$).
}
\footnotesize
\begin{tabular}{llc}
\toprule
Magnitude est. method & Phase est. method & MOS \\
\midrule
\emph{Ideal} (real magnitudes) & \emph{Ideal} (real phases) & $4.30 \pm 0.06$ \\
\emph{Ideal} (real magnitudes) & Griffin-Lim w/ $60$ iters & $3.70 \pm 0.07$ \\
\emph{Ideal} (real magnitudes) & Local Weighted Sums & $\mathbf{4.09 \pm 0.06}$  \\
Mel pseudoinverse & \emph{Ideal} (real phases) & $\mathbf{4.04 \pm 0.06}$ \\
Mel pseudoinverse & Griffin-Lim w/ $60$ iters & $2.48 \pm 0.09$ \\
Mel pseudoinverse & Local Weighted Sums & $2.51 \pm 0.09$  \\
\bottomrule
\end{tabular}
\label{tab:gl}
\end{table}

From these results, we conclude that an ideal solution to \emph{either} magnitude or phase estimation can be coupled with a good heuristic for the other to produce high-quality speech. 
While the ground truth speech is still significantly more natural than that of ideal+heuristic strategies, 
the MOS for these methods are only $2$-$9$\% worse than the ground truth ($p < 0.05$). 
Of these two problems, 
we focus on building magnitude estimation strategies as the conventional heuristic (pseudoinverse) is comparatively primitive to heuristics used for phase estimation.

As a secondary conclusion, we observe that---for our speech data---using LWS for phase estimation from real spectrograms yields significantly higher MOS than using Griffin-Lim. 
Given that it is faster \emph{and} yields significantly more natural speech, we recommend that all TTS research use LWS as a phase estimation baseline instead of Griffin-Lim. 
Henceforth, all of our experiments that require phase estimation use LWS.

\section{Adversarial vocoding}
\label{sec:methodology}
Our goal is to invert a mel spectrogram feature representation into a time domain waveform representation. 
In the previous section, we demonstrated the potential of the magnitude estimation subproblem for achieving this goal in combination with the LWS phase estimation heuristic. 
A common heuristic for magnitude estimation is performed by multiplying the mel spectrogram with the approximate inverse of the mel transformation matrix.
Since the mel spectrogram is a lossy compression of the magnitude spectrogram, a simple linear transformation is an oversimplification of the magnitude estimation problem.


In order to improve on heuristic magnitude estimation, we formulate it as a generative modeling problem and propose a Generative Adversarial Network (GAN) \cite{goodfellow2014generative} based solution.\footnote{GANs have been previously used for phase estimation~\cite{oyamada2018generative} and to enhance speech both before~\cite{kaneko2017generative} and after~\cite{tanaka2018wavecyclegan} vocoding.}
GANs are generative models which seek to learn latent structure in the distribution of data. They do this by mapping samples $\bm{z}$ from a prior distribution $p_Z$ to samples $\bm{y}$,
$G: \bm{z}\rightarrow \bm{y}$. 
For our purpose, we use a variation of GAN called \textit{conditional} GAN \cite{cGAN} to model the conditional probability distribution of magnitude spectrograms given a mel spectrogram. 
The pix2pix method~\cite{pix2pix} demonstrates that this conditioning information can be a structurally-rich image, extending GANs to learn stochastic mappings from one image domain (spectrogram domain in our case) to another.
We adapt it for our task.



The conditional GAN objective to generate appropriate magnitude spectrograms $\bm{y}$ given mel spectrograms $\bm{x}$ is:
\begin{align}
    \mathcal{L}_{\mathit{cGAN}}(G,D) = &\mathbb{E}_{\bm{x},\bm{y}}[\log D(\bm{x},\bm{y})] + \nonumber \\
                 &\mathbb{E}_{\bm{x},\bm{z}}[\log (1-D(\bm{x},G(\bm{x},\bm{z}))],\label{cGAN_equation}
\end{align}
where the generator $G$ tries to minimize this objective against an adversary $D$ that tries to maximize it. i.e $G^*  = \arg\min_G \max_D \mathcal{L}_{\mathit{cGAN}}(G,D)$. 
In such a conditional GAN setting, 
the generator tries to ``fool'' the discriminator by generating \textit{realistic} magnitude spectrograms that correspond to the conditioning mel spectrogram.  
Previous works \cite{pix2pix,segan}  have shown that it is beneficial to add a secondary component to the generator loss in order to minimize the $L_1$ distance between the generated output $G(\bm{x},\bm{z})$ and the target $\bm{y}$.
This way, the adversarial component encourages the generator to generate more realistic results,  while the $L_1$ objective ensures the generated output is close to the target.
\begin{align}
    \mathcal{L}_{L1}(G) = \mathbb{E}_{\bm{x},\bm{y},\bm{z}}[{||\bm{y}-G(\bm{x},\bm{z})||}_1].\label{L1_equation}
\end{align}
Our final objective therefore becomes:
\begin{align}
    G^*  = \arg\min_G\max_D \mathcal{L}_{\mathit{cGAN}}(G,D) + \lambda \mathcal{L}_{L1}(G).\label{full_objective}
\end{align}
Here, $\lambda$ is a hyperparameter which determines the trade-off between the $L_1$ loss and adversarial loss.


\subsection{Network architecture}

\begin{figure}[htbp]
    \centering
    \includegraphics[width=0.55\columnwidth]{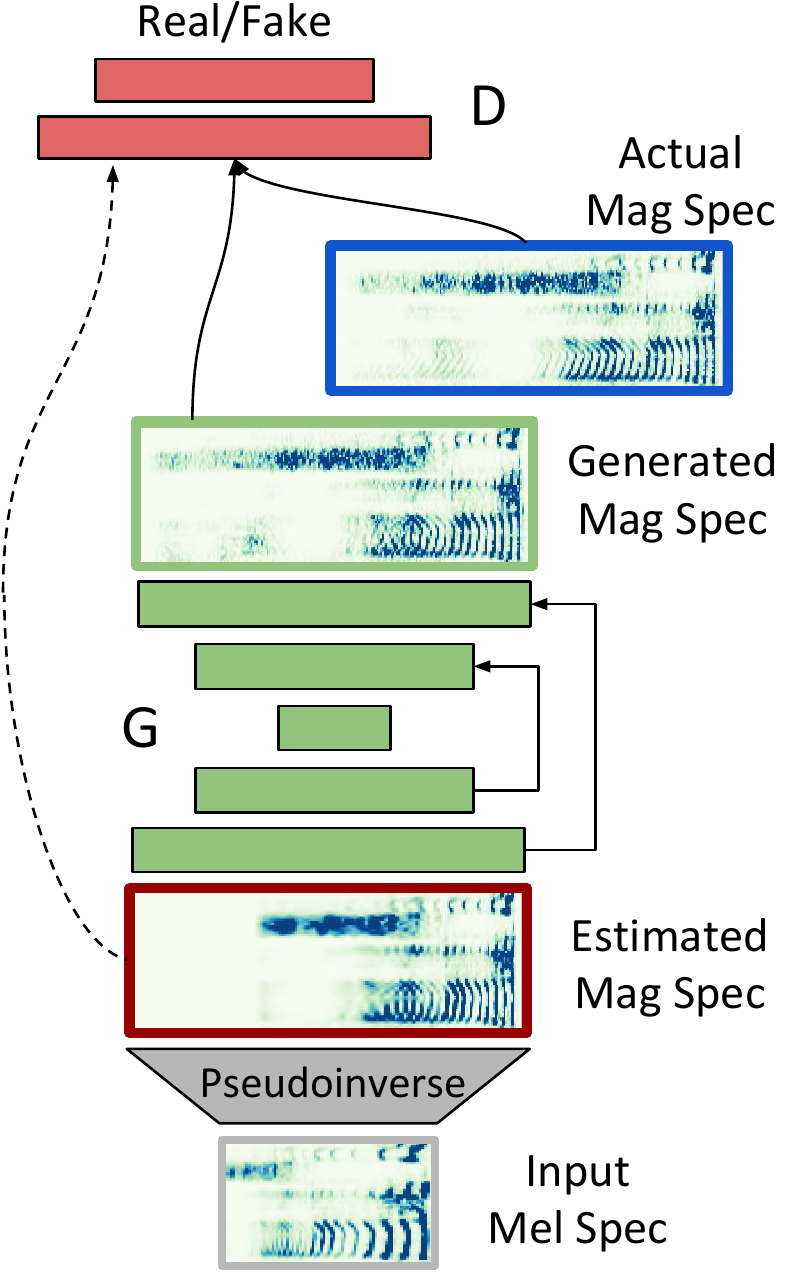}
    \caption{Adversarial Vocoder Model: The generator performs an image-to-image translation from the estimated magnitude spectrogram to the actual magnitude spectrogram guided by an adversarial loss from the discriminator and the $L_1$ distance between the generated and actual magnitude spectrogram }
    \label{fig:model}
\end{figure}

Figure \ref{fig:model} shows our setup for adversarial inversion of the mel spectrogram into a magnitude spectrogram.

\textbf{Generator}~~~The generator network $G$ takes as input the linear-amplitude mel spectrogram representation $x$ of shape $(n, \mathit{melBins})$ and generates a magnitude spectrogram of shape $(n, 513)$; $n=256$ (nearly $3$ seconds) in all of our experiments. The generator first estimates the magnitude spectrorgram through a fixed (non trainable) linear projection of the mel spectrogram using the approximate inverse of the mel transformation matrix. The estimated magnitude spectrogram goes through a convolution based encoder-decoder architecture with skip connections as in pix2pix~\cite{pix2pix}. 
Past works~\cite{Mathieu2016DeepMV,pix2pix} have noted that generators similar to our own empirically learn to ignore latent codes leading to deterministic models. 
We adopt the same policy of using dropout at both training and test time to force the model to be stochastic (as our task is not a one-to-one mapping). 
Additionally, we also train a smaller generator \textit{(Advoc - small)} with fewer convolutional layers and fewer convolutional channels. 
We omit the specifics of our architecture for brevity, however we point to our codebase (link in footnote of previous page) for precise model implementations.

\textbf{Discriminator}~~~Previous works have
found
that training generators similar to our own using just an $L_1$ or $L_2$ loss produces images with reasonable global structure (spatial relationships preserved) but poor local structure (blurry)~\cite{pathakCVPR16context,zhang2016colorful}.
As in~\cite{pix2pix}, we combine an $L_1$ loss with a discriminator which operates on \emph{patches} (subregions) of a spectrogram to help improve the ``sharpness'' of the output.
Our discriminator takes as input the estimated spectrogram and \emph{either} the generated or real magnitude spectrogram. 
Thus, in order to satisfy the discriminator, the generator needs to produce magnitude spectrograms that both correspond to the mel spectrogram \emph{and} look realistic.
To complete our adversarial vocoding pipeline, we combine generated magnitude spectrograms with LWS-estimated phase spectrograms and use the inverse STFT to synthesize audio.




\section{Experiments}
\label{sec:experiments}
We focus our primary empirical study on the publicly available LJ Speech dataset~\cite{ljspeech}, which is popularly used in TTS research~\cite{waveglow,r9y9}. 
The dataset contains $13$k short audio clips ($24$ hours) of a single speaker reading from non-fiction books.

Audio is processed using the feature extraction process described in Section \ref{sec:feature}. 
We train three models for $\mathit{melBins} \in \lbrace 20, 40, 80 \rbrace$ to study the feasibility of our technique 
for 
varying
levels of mel compression. Each of the models is trained for 100,000 mini-batch iterations using a batch size of 8 which corresponds to
12 hours of wall clock training time using a NVIDIA 1080Ti GPU.
We set the regularization parameter $\lambda = 10$ and use the Adam optimizer~\cite{kingma2014adam} ($\alpha = 0.0002$).

\subsection{Vocoding LJ Speech mel spectrograms}
\label{sec:ljspeechexp}
In this study we are concerned with vocoding both real mel spectrograms extracted from the LJ Speech dataset
\emph{and} mel spectrograms generated by a language-to-spectrogram model~\cite{shen2018natural} trained on LJ Speech. 
We compare both our large (\emph{AdVoc}) and small (\emph{AdVoc-small}) adversarial vocoder models to 
the mel pseudoinverse magnitude estimation heuristic combined with LWS (\emph{Pseudoinverse}),
a \emph{WaveNet} vocoder~\cite{shen2018natural},  
and the recent \emph{WaveGlow}~\cite{waveglow} method. 
We cannot directly compare to the \emph{Parallel WaveNet} approach because it is an end-to-end TTS method rather than a vocoder~\cite{oord2017parallel}.

We randomly select $100$ examples from the holdout dataset of LJ Speech and convert them to mel spectrograms. 
We also synthesize mel spectrograms for each transcript of these same examples using the language-to-spectrogram module from Tacotron 2~\cite{shen2018natural}.
We vocode both the real and synthetic spectrograms to audio using the five methods outlined in the previous paragraph. 
Audio from each method can be found in our sound examples (footnote of first page).

To gauge the relative quality of our methods against others,
we conduct two mean opinion score (comparative) studies with human judges on Amazon Mechanical Turk. 
In the first user study, 
judges evaluate a batch of six versions of the same utterance: the original utterance and the spectrogram of that utterance vocoded by the five aforementioned methods. 
In the second user study, 
we show each judge a batch consisting of the real utterance and five vocodings of a synthetic spectrogram with the same transcript. 
In all user studies, the ordering of the waveforms is randomized in each batch but the waveforms in a batch always pertain to the same utterance. 
Judges are asked to rate the naturalness of each on a subjective $1$--$5$ scale with $1$ point increments. 
Each batch is reviewed by $8$ different reviewers resulting in $800$ evaluations of each strategy.
We display 
mean opinion scores
in Table~\ref{tab:gl}. 
We also include the speed of each method (relative to real time) as measured on GPU, and the sizes of each model's parameters in megabytes. 


\begin{table}[t]
\centering
\caption{Comparison of vocoding methods on mel spectrograms with $80$ bins. We display comparative mean opinion scores from two separate user studies for vocoding spectrograms extracted from real speech (MOS-Real) and spectrograms generated by a state-of-the-art TTS method (MOS-TTS) with $95$\% confidence intervals. $\mathbf{\times}$ RT denotes the speed up over real time; higher is faster. MB denotes the size of each model in megabytes.
}
\footnotesize
\begin{tabular}{lcccr}
\toprule
Source & MOS-Real & MOS-TTS & $\times$ RT & MB \\
\midrule
Real data & $4.16 \pm 0.06$& $4.28 \pm 0.07$  & $1.000$ &   \\
Pseudoinverse & $2.91 \pm 0.10$ & $2.12 \pm 0.09$ & $8.836$ & $0.2$ \\
WaveNet \cite{oord2016wavenet} & $3.98 \pm 0.07$ & $3.87 \pm 0.07$ & $0.003$ & $95.0$ \\
WaveGlow \cite{waveglow} & $4.09 \pm 0.06$ & $3.89 \pm 0.07$ & $1.229$ & $334.7$ \\
AdVoc & $3.78 \pm 0.07$ & $2.91 \pm 0.08$ & $3.111$ & $207.7$ \\
AdVoc-small & $3.68 \pm 0.07$ & $3.09 \pm 0.07$ & $3.437$ & $16.0$ \\
\bottomrule
\end{tabular}
\label{tab:realMOS}

\end{table}

Our results demonstrate that---for both real \emph{and} synthetic spectrograms---our adversarial magnitude estimation technique (AdVoc) significantly outperforms magnitude estimation using the pseudoinverse of the mel basis.
Our method is more than $1000 \times$ faster than the autoregressive WaveNet vocoder and $2.5 \times$ faster than WaveGlow vocoder.

\begin{table}[t]
\centering
\caption{Comparison of heuristic and adversarial vocoding of spectrograms with different levels of mel compression. Adversarial vocoding can vocode highly compressed mel spectrograms with relatively less drop in speech naturalness as compared to a heuristic.}
\footnotesize
\begin{tabular}{lcc}
\toprule
Source & $melBins$ & MOS \\
\midrule
Real data & & $4.05 \pm 0.07$ \\
Pseudoinverse & $20$ & $2.68 \pm 0.10$ \\
Pseudoinverse & $40$ & $2.84 \pm 0.10$ \\
Pseudoinverse & $80$ & $3.25 \pm 0.09$ \\
AdVoc & $20$ & $3.75 \pm 0.07$ \\
AdVoc & $40$ & $3.79 \pm 0.07$ \\
AdVoc & $80$ & $3.86 \pm 0.07$ \\
\bottomrule
\end{tabular}
\label{tab:compression}
\end{table}

Additionally, we train our models to perform magnitude estimation on representations with higher compression. 
Specifically, we train our model to vocode mel spectrograms with $20$, $40$ and $80$ bins. 
We compare our adversarial magnitude estimation method against magnitude estimation using the pseudoinverse of the mel basis. 
We conduct a comparative user study using the same methodology as previously outlined. 
Our results in 
Table~\ref{tab:compression} demonstrate that our model can vocode highly compressed mel spectrogram representations with relatively little drop in the perceived audio quality as compared to the pseudoinversion baseline (audio examples in footnote of first page).


\subsection{Unsupervised audio synthesis}

\begingroup
\setlength{\tabcolsep}{4pt}
\begin{table}[t]
\centering
\caption{Combining our adversarial vocoding approach with GAN-generated mel spectrograms outperforms our prior work in unsupervised generation of individual words by all metrics.}
\vspace{2mm}
\footnotesize
\begin{tabular}{l|c|cc}
\multicolumn{1}{c}{} & \multicolumn{1}{c}{\emph{Quantitative}} & \multicolumn{2}{c}{\emph{Qualitative}} \\
\toprule
Source & Inception score & Acc. & MOS  \\
\midrule
Real data & $8.01 \pm 0.24$ & $0.95$ & $3.9 \pm 0.15$ \\
WaveGAN \cite{donahue2019wavegan} & $4.67 \pm 0.01$ & $0.58$ & $2.3 \pm 0.18$   \\
SpecGAN  \cite{donahue2019wavegan} + Griffin-Lim & $6.03 \pm 0.04$ & $0.66$ & $1.9 \pm 0.17$ \\
MelSpecGAN + AdVoc & $6.63 \pm 0.03$ & $0.71$ & $3.4 \pm 0.20$ \\
\bottomrule
\end{tabular} 
\label{tab:sc09Small}
\end{table}
\endgroup

In this section we are concerned with the \emph{unsupervised} generation of speech (as opposed to supervised generation in the case of TTS). 
We focus on the SC09 digit generation task proposed in our previous work~\cite{donahue2019wavegan}, 
where the goal is to learn to generate examples of spoken digits ``zero'' through ``nine'' \emph{without} labels. 
We first train a GAN to generate mel spectrograms of spoken digits (\emph{MelSpecGAN}), then train an adversarial vocoder to generate audio conditioned on those spectrograms. 
Using a pretrained digit classifier, 
we calculate an Inception score~\cite{salimans2016improved} for our approach, finding it to outperform our previous state-of-the-art results by $9$\%. 
We also calculate an ``accuracy'' by comparing human labelings to classifier labels for our generated digits,
finding that our adversarial vocoding-based method outperforms our previous results (Table~\ref{tab:sc09Small}).

\section{Conclusion}

In this work we have shown that solutions to \emph{either} the magnitude or phase estimation subproblems within common vocoding pipelines can result in high-quality speech. 
We have demonstrated a learning-based method for magnitude estimation which significantly improves upon popular heuristics for this task.
We demonstrate that our method can integrate with an existing TTS pipeline to provide comparatively fast waveform synthesis. 
Additionally, our method has advanced the state of the art in unsupervised small-vocabulary speech generation.

\section{Acknowledgements}

The authors would like to thank Bo Li for helpful discussions about this work. 
This research was supported by the UC San Diego Chancellor’s Research Excellence Scholarship program. 
Thanks to NVIDIA for GPU donations which were used in the preparation of this work.

\bibliographystyle{IEEEtran}

\bibliography{advoc}


\end{document}